\documentclass[journal]{IEEEtran}
\usepackage{graphicx}
\usepackage{multirow}
\usepackage{subfigure}
\usepackage{amsmath}
\usepackage{array}

\usepackage[belowskip=-15pt,aboveskip=0pt]{caption}

\ifCLASSINFOpdf

\else

\fi

\begin{document}
%
\title{Analysis and Modeling Experiment Performance Parameters of Routing Protocols in MANETs and VANETs}

\author{\IEEEauthorblockN{S. Sagar, N. Javaid, Z. A. Khan$^{\S}$, J. Saqib, A. Bibi, S. H. Bouk} \\
                Department of Electrical Engineering, COMSATS\\ Institute of
                Information Technology, Islamabad, Pakistan. \\
                $^{\S}$Faculty of Engineering, Dalhousie University, Halifax, Canada.
             }

\maketitle

\begin{abstract}
In this paper, a framework for experimental parameters in which Packet Delivery Ratio (PDR), effect of link duration over End-to-End Delay (E2ED) and Normalized Routing Overhead (NRO) in terms of control packets is analyzed and modeled for Mobile Ad-Hoc NETworks (MANETs) and Vehicular Ad-Hoc NETworks (VANETs) with the assumption that nodes (vehicles) are sparsely moving in two different road. Moreover, this paper contributes the performance comparison of one Proactive Routing Protocol; Destination Sequenced Distance vector (DSDV) and two reactive protocols; DYnamic Source Routing (DSR) and DYnamic MANET On-Demand (DYMO). A novel contribution of this work is enhancements in default versions of selected routing protocols. Three performance parameters; PDR, E2ED and NRO with varying scalabilities are measured to analyze the performance of selected routing protocols with their original and enhanced versions. From extensive simulations, it is observed that DSR outperforms among all three protocols at the cost of delay. NS-2 simulator is used for simulation with TwoRayGround propagation model to evaluate analytical results.
\end{abstract}


\IEEEpeerreviewmaketitle
\section{Background}



Mobile Ad-Hoc NETwork (MANET) is a self-configuring network of mobile nodes connected with wireless link in which each mobile acts as specialized router, thus, it is capable of forwarding packets to other nodes. Vehicular Ad-Hoc NETwork (VANET) is a special type of MANET used to provide communication between vehicles moving with high mobility.

In wireless ad-hoc networks, routing protocols are used to calculate efficient routes. These protocols are divided into two main categories with respect to their routing behavior; on-demand (reactive) and table driven (proactive). Reactive routing protocols calculate routes for destination in the network, when it is needed therefore these are known as on-demand routing protocols. Proactive protocols are based on periodic exchange of control messages and maintaining routing tables, that is why these are known as table-driven routing protocols for complete implementation of topology locally. Reactive protocols usually takes more time to find a route as compared to a proactive protocol. For our analysis, we have selected two reactive routing protocols, DSR [1] and DYMO [2] and one proactive routing protocol DSDV [3]. Moreover, we also enhance DSR and DYMO to obtain efficient performance. To validate the efficiency of these enhancements, simulations are performed in NS-2 by considering different scalabilities using RandomWay Point propagation model.


\section{Related Work and Motivation}
Several studies have been made for comparing different MANETs routing protocols using different performance metrics. Performance study which is presented in [4], discusses a delay time analysis for multi-hop Vehicle to Vehicle (V2V) communication over linear VANETs. Authors in this paper discuss only about Packet Delivery rate (PDR) and End-to-End Delay (E2ED), however, we have also discussed about the Normalized Routing Overhead (NRO).

Performance analysis of two reactive protocols, AODV and DSR is compared by A. Shastri \textit{et al.} [5] with varying pause time, scalability and number of connections only in VANETs. On the other hand, we compare reactive protocols with proactive ones, like AODV, AOMDV, DSDV and DYMO are evaluated by Mohammad Azouqa \textit{et al.} [6] with performance metrics PDR, AE2ED and NRO versus number of nodes in VANETs.

Performance evaluation of AODV and DSR with varying pause time and node density over TCP and CBR connection in VANETs is compared by [7].

Saishree Bharadwaj.P. \textit{et al.} in [8], compare the performance of AODV and DSDV in Urban Scenario of VANETs.

Rajeshwar Singh \textit{et al.} [9] evaluate the performance of DSDV and DSR using performance metrics; throughput and Packet Delivery Fraction (PDF) with varying scalability in MANETs.

In [10], authors compared AODV, DSR and DSDV on the basis of TCP traffic pattern only in MANETs.



DYMO is a reactive routing protocol and the main candidate for the upcoming reactive MANET routing protocols. It is based on the work and experience from previous reactive routing protocols, especially AODV and DSR [11].



The studies that have been done so far from [4] to [8], compare the performance of routing protocols in VANETs only and
the studies from [9] to [11] compare the performance of protocols only in MANETs. In this paper, we compare two reactive protocols; DSR and DYMO and a proactive protocol; DSDV in both MANETs and VANETs. A novel contribution of this work is enhancement of DSR and DYMO protocols to improve the efficiency.

\section{Modeling Experiment Performance Parameters}

In [4], author derives the equation for Average End-to-End Delay (AE2ED) and PDR by using probability distribution with some assumptions. One of the assumption is that the probability of a segment $x$ of a single road contains the car is $\lambda x$ at any time $t$. The initial inter-vehicle spacing is $d\{0\}$ and $R$ is the range of node. The Probability of First Time communication between two nodes (vehicles) is:

\begin{align}
Pr(d{0} < R) = \lambda e^{-\lambda x}
\end{align}

 \subsection{Packet Delivery Ratio, $\rho_0$}
From the assumption of [4], the probability that the road segment $x$ of one direction contains the node/vehicle is $\lambda x$ and the probability of First Time communication between two nodes (vehicles) in eq. (1). If the inter-vehicle spacing $d\{0\} < R$ between any two nodes/vehicles then communication is easily possible and the probability is $\int_0^R \lambda
e^{-\lambda x}\,\mathrm{d}x = 1-e^{-\lambda R}$. If $d\{0\} > R$, the probability of one step wait time is
$\gamma\left(\frac{x-R}{t^2}\right).p\left(\frac{x-R}{t}\right)$. The PDR: $\rho_0$ is

\begin{align}
\begin{split}
\rho_0 = 1-e^{-\lambda R} + \int_R^\infty \lambda e^{-\lambda x}\ \\ \int_0^T \gamma \left(\frac{x-R}{t^2}\right)p\left(\frac{x-R}{t}\right) \mathrm{d}t\mathrm{d}x
\end{split}
\end{align}

\begin{align}
\begin{split}
 = \lambda \gamma \int_0^\infty  e^{-\lambda (x+R)} (1- q\left(\frac{x}{T}\right))\mathrm{d}x
\end{split}
\end{align}\newline where, $p$ is the probability distribution function of velocity: $p(v)$ which is:

\begin{align}
p(v) = \frac{1}{\sigma \sqrt{2\pi}}e^{-\frac{v^2}{2\sigma^2}}
\end{align}\newline where, $\sigma$ is variance and

\begin{align}
q(v) = \int_{-\infty}^v  p(u) \mathrm{d}u
\end{align}

We assume a road with two different directions $D_1$ and $D_{1}'$ or $D_2$ and $D_{2}'$, then the probability that the road segment $x$ contains the node (vehicle) in any of the direction, as shown in Fig. 1 is $\lambda^2 x^2 + 2\lambda x$ and the probability of First Time communication is:

\begin{align}
Pr(d\{0\} < R) = \lambda e^{-\lambda^2 x^2 + 2\lambda x}
\end{align}

So, the probability that inter vehicle spacing $d\{0\}< R$ is $\int_0^R \lambda e^{-(\lambda^2 x^2 + 2 \lambda x)}\ \mathrm{d}x$ and $\rho_0$ is

\begin{align}
\begin{split}
\rho_0 = \int_0^R \lambda e^{-(\lambda^2 x^2 + 2 \lambda x)}\ \mathrm{d}x + \int_R^\infty \lambda e^{-(\lambda^2 x^2 + 2 \lambda x)}\ \\ \int_0^T \gamma\left(\frac{x-R}{t^2}\right)p\left(\frac{x-R}{t}\right) \mathrm{d}t\mathrm{d}x
\end{split}
\end{align}

\begin{align}
\begin{split}
= \lambda \gamma \int_0^\infty  e^{-(\lambda^2 (x+R)^2 + 2 \lambda (x+R))} (1- q\left(\frac{x}{T}\right))\mathrm{d}x
\end{split}
\end{align}\newline where, $T$ is period during which nodes(vehicles) wait for communication.

\begin{figure}[h]
\begin{center}
\includegraphics[height=6 cm, width=8 cm]{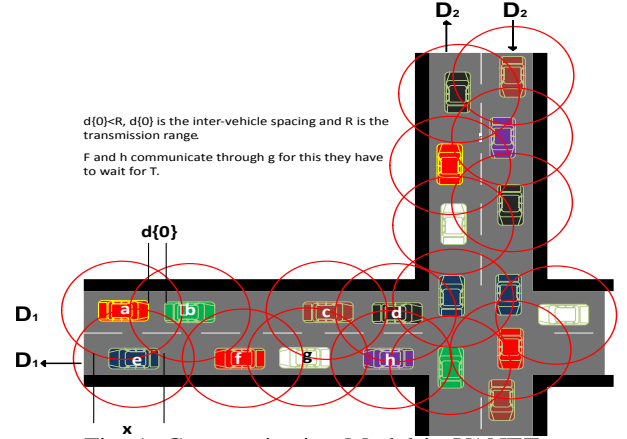}
\end{center}
\vspace{-0.4cm}
\caption{Communication Model in VANETs}
\vspace{0.2cm}
\end{figure}
\vspace{-0.2cm}
\subsection{Average End-to-End Delay, $\tau_0$}

$\tau_0$ is also calculated by using one step wait time and probability given in eq.\ (1) which is given by [4]. So,

\begin{align}
\tau_0 &= \int_R^\infty \lambda e^{-\lambda x}\ \int_0^T t . \gamma \left(\frac{x-R}{t^2}\right).p\left(\frac{x-R}{t}\right) \mathrm{d}t \mathrm{d}x
\\ &= \lambda \gamma e^{-\lambda R} \int_0^\infty e^{-\lambda x}\ \int_0^T \left(\frac{x}{t}\right).p\left(\frac{x}{t}\right)) \mathrm{d}t \mathrm{d}x
\end{align}\newline where,

\begin{align}
\int_0^T \left(\frac{x}{t}\right).p\left(\frac{x}{t}\right) \mathrm{d}t = \frac{-x. E_i(- \frac{x^2}{2\sigma^2 T^2})}{2\sigma \sqrt {2\pi}}
\end{align}

$\tau_0$ in terms of $\beta$ is

\begin{align}
\tau_0 = \frac{\lambda \gamma T^2 e^{-\lambda R}}{\sqrt{2\pi}} . \beta(\lambda T \sigma \sqrt{2})
\end{align}\newline while, $\beta$ is

\vspace{0.3cm}
\[\beta (z) = - \int_0^\infty xe^{-zx}Ei(-x^2)\]

and $Ei(x)$ is exponential integral which is
\begin{align}
Ei(x) = \int_{-\infty}^x \frac{e^t}{t} \mathrm{d}t
\end{align}

Similarily, according to our assumption, we can write $\tau_0$ in terms of eq. (5) as,

\begin{align}
\begin{split}
\tau_0 &= \int_R^\infty \lambda e^{-(\lambda^2 x^2 + 2 \lambda x)}\ \int_0^T t . \gamma\left(\frac{x-R}{t^2}\right). p\left(\frac{x-R}{t}\right) \mathrm{d}t \mathrm{d}x
\end{split}
\end{align}

\begin{align}
\begin{split}
& = \lambda \gamma e^{-(\lambda^2 R^2 + 2 \lambda R)} \int_0^\infty e^{-(\lambda^2 x^2 + 2 \lambda x)}\ \int_0^T \left(\frac{x}{t}\right).p\left(\frac{x}{t}\right)) \mathrm{d}t \mathrm{d}x
\end{split}
\end{align}
\vspace{-0.7cm}

\subsection{Normalized Routing Overhead, $NRO$}













\subsubsection{\textbf {DYMO}: $NRO_{DYMO}$}

For calculating NRO of DYMO, first we calculate NRO for Route Discovery (RD) and for Route Maintenance (RM). As $NRO^{DYMO}_{Total}$ is:

\begin{align}
\begin{split}
NRO^{DYMO}_{Total} = NRO^{DYMO}_{RD} + NRO^{DYMO}_{RM}
\end{split}
\end{align}

\begin{align}
\begin{split}
NRO^{DYMO}_{RD} = No: of Sources.\int_0^{TTL(R)} \lambda e^{\frac{\lambda h r}{s}}\mathrm{d}h
\end{split}
\end{align}

\begin{align}
\begin{split}
NRO^{DYMO}_{RM} = \frac{No: of Sources}{s}.[\frac{t.h}{H_{int}} + RERR_{pkts}]
\end{split}
\end{align}

While $RERR_{pkts}$ represents the number of Route ERRor (RERR) messages;

\begin{align}
\begin{split}
RERR_{pkts} = \frac{t.h}{LB_{int}}
\end{split}
\end{align}\newline where, $'h'$ is hop-count, $'r'$ shows the routing packet, $'s'$ show the generated packets, $TTL(R)$ shows the $TTL$ value of the ring $R$ during Expanding Ring Search (ERS), $No: of Sources$ are $12$, $LB_{int}$ is the link breakage occurance, $'H_{int}'$ is periodic update time for link sensing i.e., $1\ sec$ in case of DYMO routing protocol and $t$ is the time period.

\subsubsection{\textbf {DSR}: $NRO_{DSR}$}

Also equation for NRO of DSR is same as for DYMO but link monitering in DSR is done on MAC layer. So,
\vspace{-0.1cm}
\begin{align}
\begin{split}
NRO^{DSR}_{Total} = NRO^{DSR}_{RD}
\end{split}
\end{align}
\vspace{-0.2cm}
\begin{align}
\begin{split}
NRO^{DSR}_{RD} = No: of Sources.\int_0^{TTL(R)} \lambda e^{\frac{\lambda h r}{s}}\mathrm{d}h
\end{split}
\end{align}

DSR uses Packet Salvaging (PS) technique to maintain the routes when any link breakage occurs.



\vspace{0.3cm}
\subsubsection{\textbf {DSDV}: $NRO_{DSDV}$}

Now for Proactive Protocols DSDV, formula for NRO is

\begin{align}
\begin{split}
NRO^{DSDV}_{Total} = NRO^{DSDV}_{RU_{Per}} + NRO^{DSDV}_{RU_{trig}}
\end{split}
\end{align}

\begin{eqnarray}
NRO=
  \begin{cases}
   \frac{No: of Sources}{s}[\frac{t.h}{H_{int}} & for \ NRO_{RU_{PU}}] \\
   \frac{No: of Sources}{s}[\frac{t.h}{trig_{int}} & for \ NRO_{RU_{trig}}] \\
   \end{cases}
\end{eqnarray}




Here the periodic update interval $H_{int}$ is $15\ sec$ and $trig_{int}$ depends on the event of breakage. As MAC layer notify this breakage after $0.8sec$. Therefore, we have taken it as $0.8\ sec$ for active roues.
The complete information about NRO of DSR and DYMO is discussed in [12] and information about NRO of DSDV is given in [13].







\subsection{Numerical Results and Graphs}

From [14], the rate parameter $\lambda (veh/sec)$ is approximated as:
\vspace{-0.1cm}
\begin{align}
\begin{split}
\lambda \approx \frac{T_v}{3600\ V}
\end{split}
\end{align}

Here, $V$ is the speed in $m/sec$ which is $11.11m/sec$ and $T_v$ is traffic volume which is in $veh/hour$ which is 10 for midnight and 70 for morning.

In this paper, we measure different values of $\lambda$ from eq. (24), in which one is $\lambda = 0.00025 (veh/sec)$ at midnight when there are less nodes(vehicles) whereas in case of morning $\lambda = 0.00175$. Wait Time $T = 1,\ 2,\ 3,\ 4,\ 5,\ 10,\ 15,\ 20,\ 25,\ sec$. Also the value of $\gamma$ is approximately equal to one, transmission range of our model is $R = 250m$ and the average transmission time $\delta = 300msec$. The average delay is $\tau_0 + \delta$ and PDR is $\rho_0$. Fig. 1 shows the graphs of $\rho_0$ and $\tau_0$.

From Fig. 2(a), we observe that dependency of $\rho_0$ on $T$ is small. Variation in $\rho_0$ is very small in both midnight and morning varying $T$. Similarly, in Fig. 2(b), when the value of $T$ is small $(T \leq 10\ sec)$, delay in both midnight and in morning is same and delay increases with increase in $T$.

\begin{figure}[!h]
  \centering
 \subfigure[Packet Delivery Ratio]{\includegraphics[height=2.7 cm,width=4.3 cm]{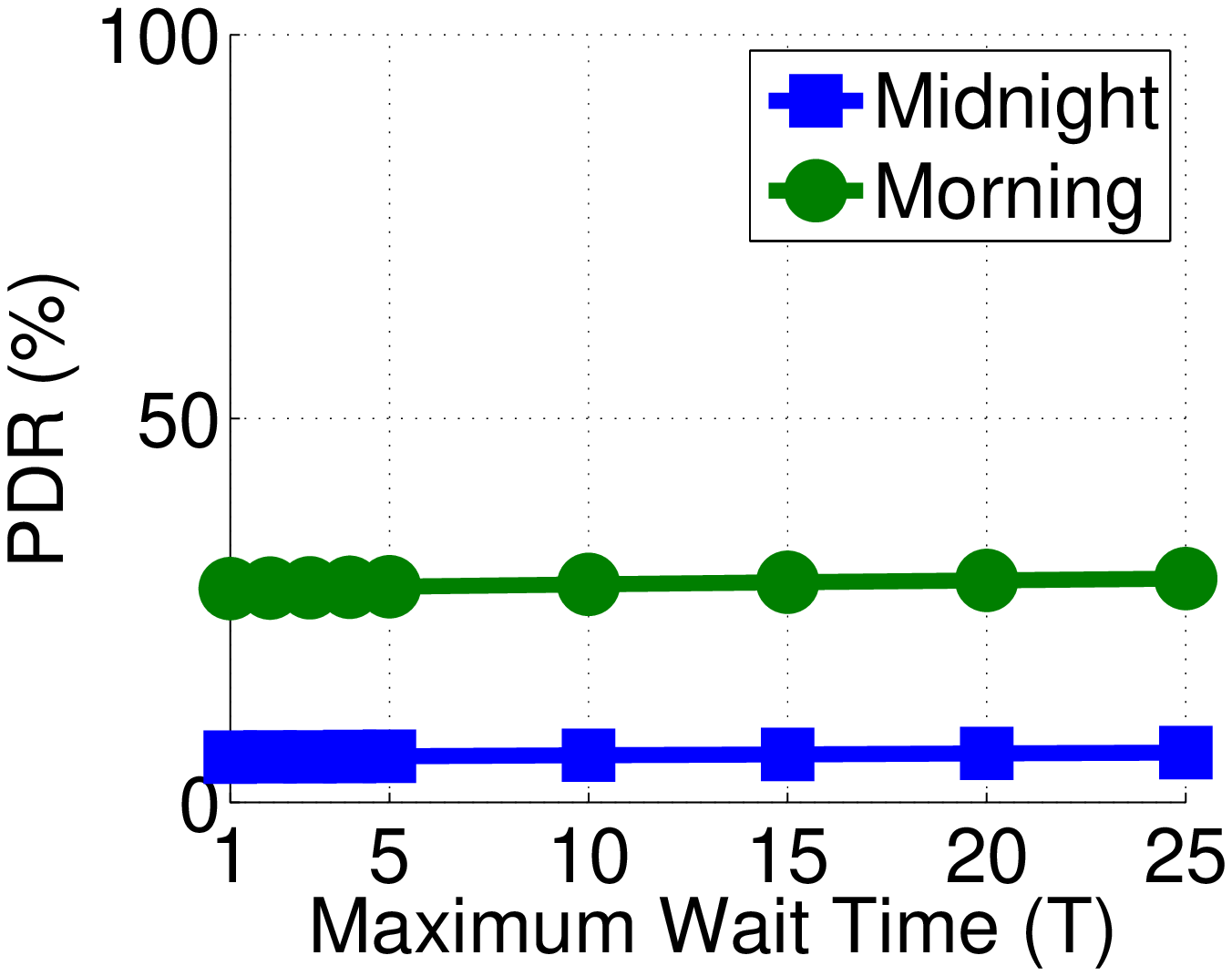}}
 \subfigure[Average Delay]{\includegraphics[height=2.7 cm,width=4.3 cm]{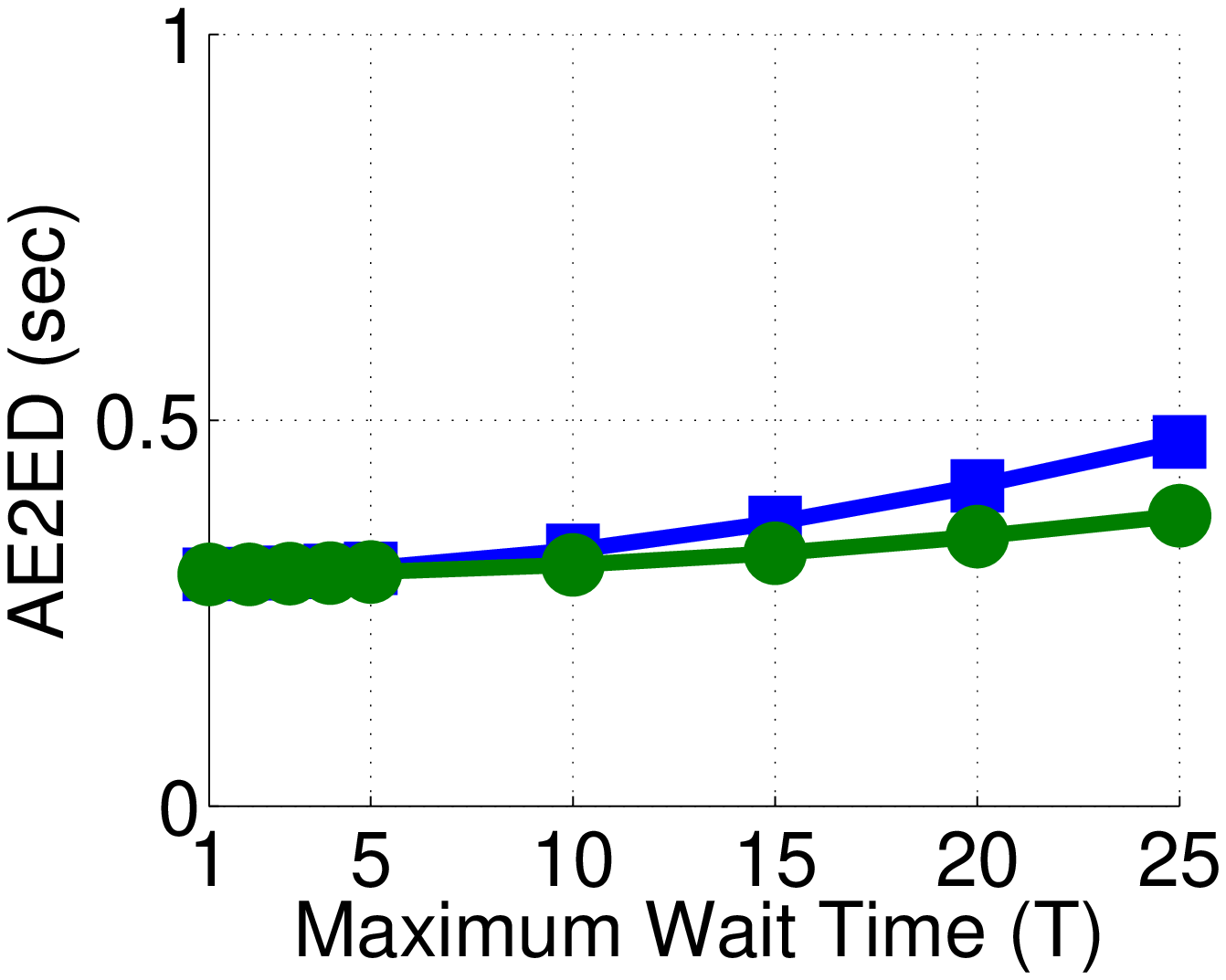}}
 \caption{Packet Delivery Ratio and Average Delay}
\end{figure}

For NRO, when we assume $10$ number of nodes $\lambda = 0.00025 (veh/sec)$. For this we have $r=4453$ and $s=16069$ in case of DYMO, for DSR $r=917$, and $s=16410$ and for DSDV $r=547$ and $s=20996$ while assuming 70-nodes, we have $r=3140$ and $s=1584$ for DYMO, $r=24692$ and $s=36146$ for DSR and $r=7178$ and $s=19223$ for DSDV. The maximum hop-count $(h = 8)$ in midnight while in morning maximum hop count $(h = 68)$ and the minimum hop count is $(h = 2)$ both in midnight and in morning.

From Fig. 3, by increasing the period $t$, NRO of all three selected paper increases but NRO of DYMO and DSR is low as compared to DSDV in both cases when $(h = 2)$ and in $(h = 8)$. When $(h = 8)$ NRO of DSDV increases very fast as compared to $(h = 2)$. Similarly, in Fig. 4, by increasing $t$, NRO of all three selected protocols is also increased.

\begin{figure}[!h]
  \centering
 \subfigure[NRO Produced with h=2]{\includegraphics[height=2.7 cm,width=4.3 cm]{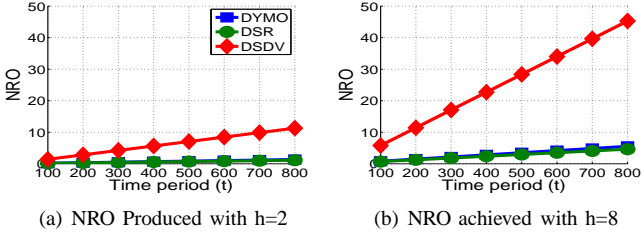}}
 \subfigure[NRO achieved with h=8]{\includegraphics[height=2.7  cm,width=4.3 cm]{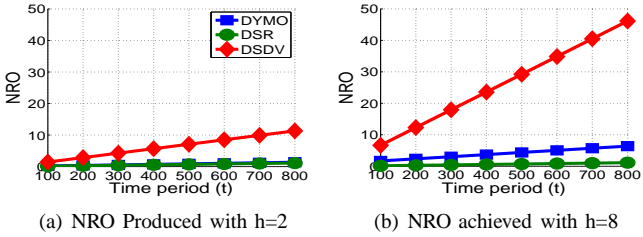}}
 \caption{NRO by protocols with varying time for 10-nodes}
\end{figure}

\begin{figure}[!h]
  \centering
 \subfigure[NRO Produced with h=2]{\includegraphics[height=2.7 cm,width=4.3 cm]{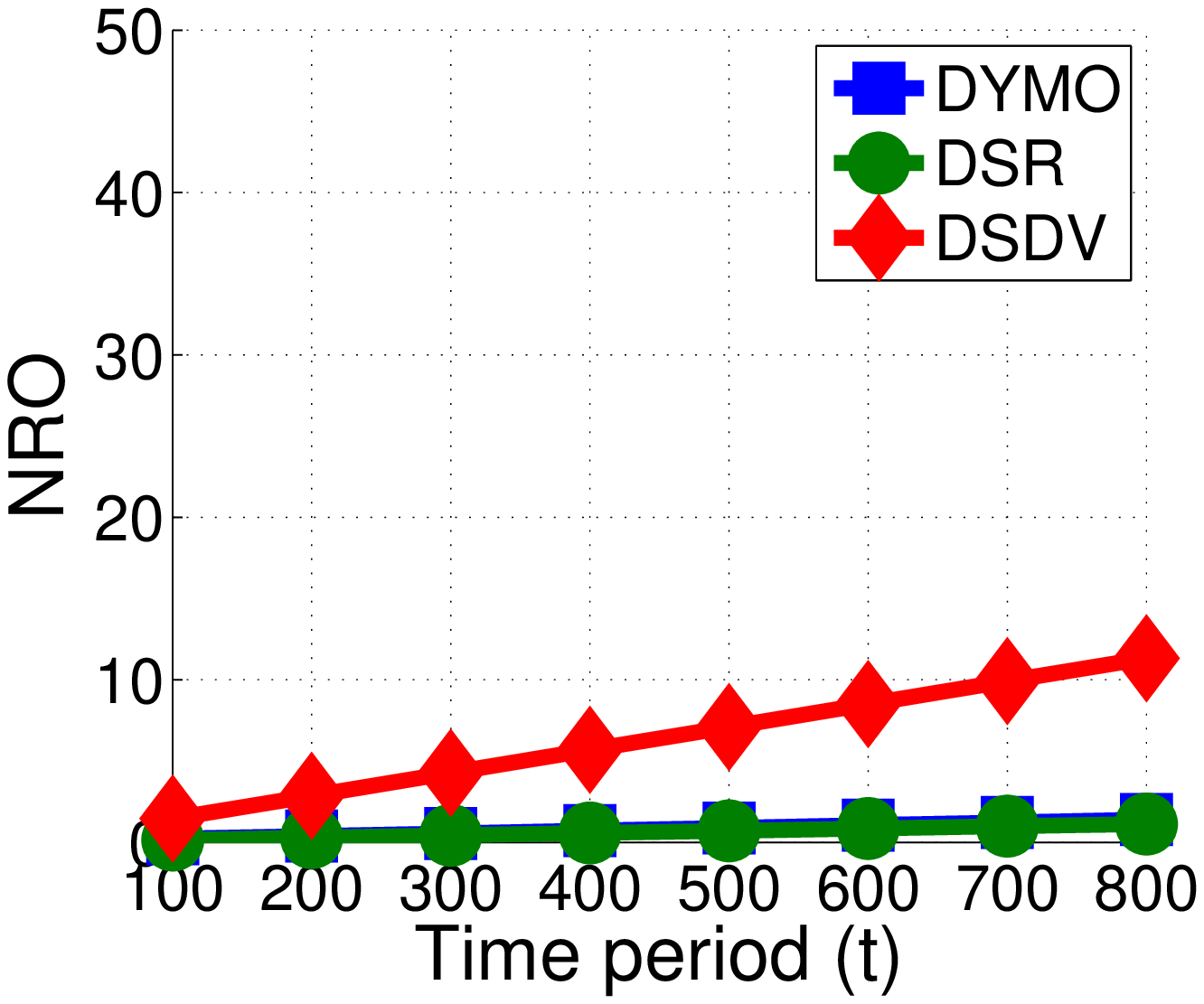}}
 \subfigure[NRO achieved with h=8]{\includegraphics[height=2.7  cm,width=4.3 cm]{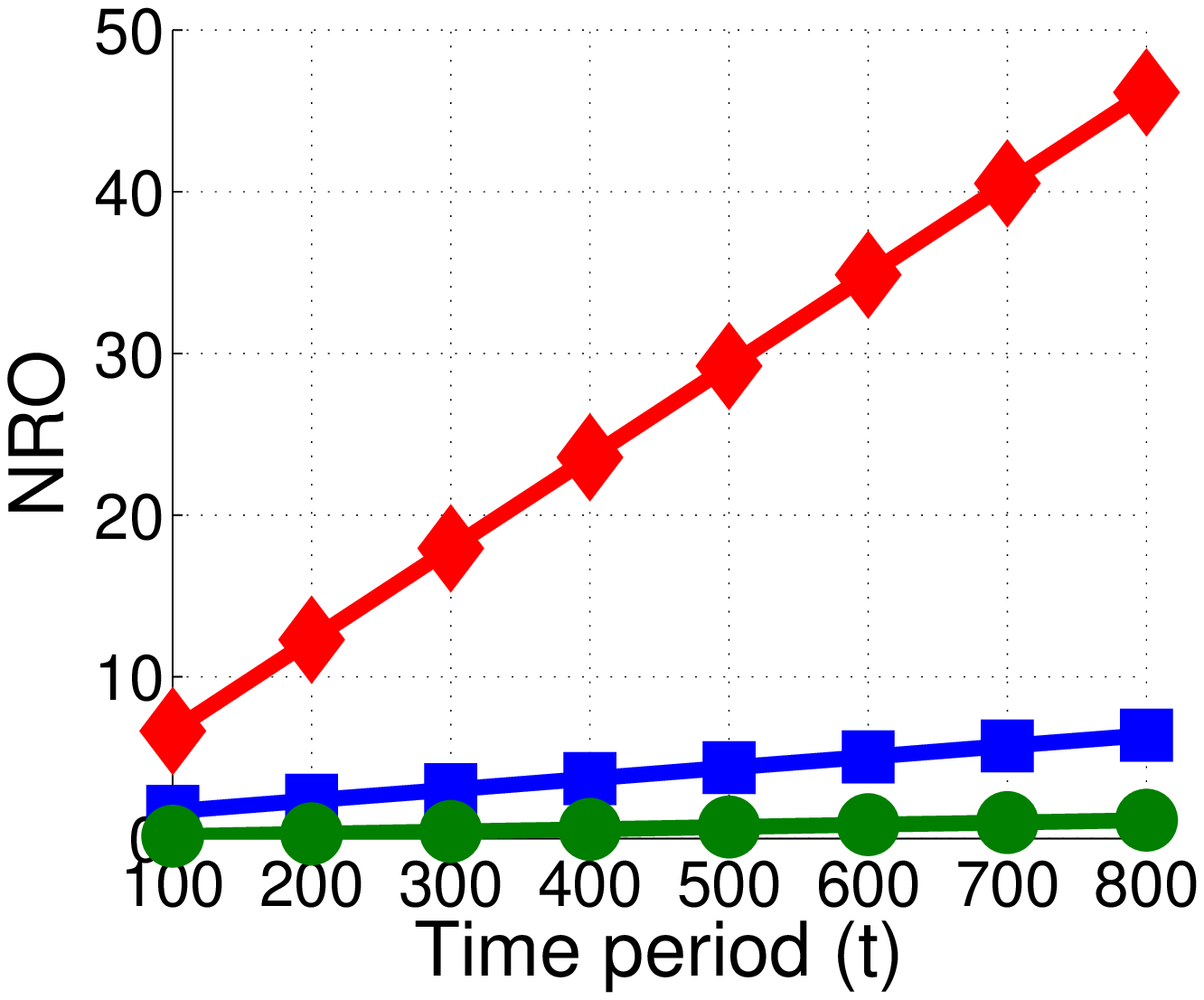}}
 \caption{NRO by protocols with varying time for 10-nodes}
\end{figure}

\section{Simulations and Discussions}
In this section, we provide the details for the simulation conducted for this study.

\vspace{0.5cm}
\begin{table}[!h]
\caption {Simulation Parameters for MANETs and VANETs}
\begin {center}
\begin{tabular}{|c|c|}
\hline
\textbf{PARAMETERS} & \textbf{VALUES}\\
\hline

NS-2 Version&	2.34\\
\hline

DYMO Implementation	&DYMOUM [15] \\
\hline

Number of nodes&	10, 20, 30,……., 70 \\
 \hline

Speed&	Uniform 40 kph\\
\hline

Data Type&	CBR\\
\hline

Simulation Time&	900 seconds\\
\hline

Data Packet Size&	1000 bytes\\
\hline

PHY Standard&	802.11/802.11p\\
\hline

Radio Propagation Model&	TwoRayGround\\
\hline

SUMO Version&	0.13\\
\hline

\end{tabular}
\end{center}
\end{table}

\vspace{-0.4cm}

We enhanced DSR and DYMO protocols. In DEF-DSR, $SEND\_BUF\_SIZE$ is 64 and $TAP\_CACHE\_SIZE$ is 1024, while in (MOD-DSR) we double $SEND\_BUF\_SIZE$ and reduce $TAP\_CACHE\_SIZE$ to one fourth. For enhancements in DYMO, $TTL\_NET\_DIAMETER$ $=10$ in DEF-DYMO is set to $30$ and  $RREQ\_WAIT\_TIME$ (= 1000 ms in DEF-DYMO) is modified by setting its value to $600 ms$.

Fig. 5 shows the percentage of PDR, AE2ED is compared in Fig. 6 and Fig. 7 show NRO against varying scalabilities.

\vspace{-0.3cm}
\subsection{PDR}
IEEE $802.11p$ MAC uses the Enhanced Distributed Channel Access (EDCA) mechanism originally provided by IEEE $802.11e$. Therefore, successful packet delivery rate of all protocols is better in VANETs as compared to MANETs, as shown in Fig. 5. Fig. 5(a)(c) depicts that PDR of all protocols is more in medium scalabilities and less in higher scalabilities, because congested networks suffer more interferences which augment drop rates. Among all selected protocols performance of DSR is high as compared to DYMO (DEF-DYMO and MOD-DYMO) and DSDV in both in MANETs and in VANETs, as depicted in Fig. 5. Incremental updates due to link breakages along with route settling time make DSDV more convergent. The reason behind such behavior of DYMO is the absence of any supplementary mechanisms during route discovery and route maintenance in DYMO. In DSR, promiscuous listening mode permits to cache multiple routes in route cache. These cached routes provide already calculated routes during RD (Route Caching) and grant alternative routes during RM (Packet Salvaging). Consequently, DSR achieves highest throughput due to quick route discovery and quick repair.

Reduction in $RREQ\_WAIT\_TIME$ and increase in $TTL\_NET\_DIAMETER$ formulate MOD-DYMO to generate less drop rates. Moreover, by increasing $SEND\_BUF\_SIZE$ and reduction of $TAP\_CACHE\_SIZE$ in MOD-DSR provide fresher routes in route cache, and thus augments PDR in both MANETs and in VANETs, as shown in Fig. 5(b)(d) comparative to Fig. 5(a)(c).

\begin{figure}[!h]
  \centering
 \subfigure[PDR of Orig. Prot.s MANETs]{\includegraphics[height=2.7 cm,width=4.3 cm]{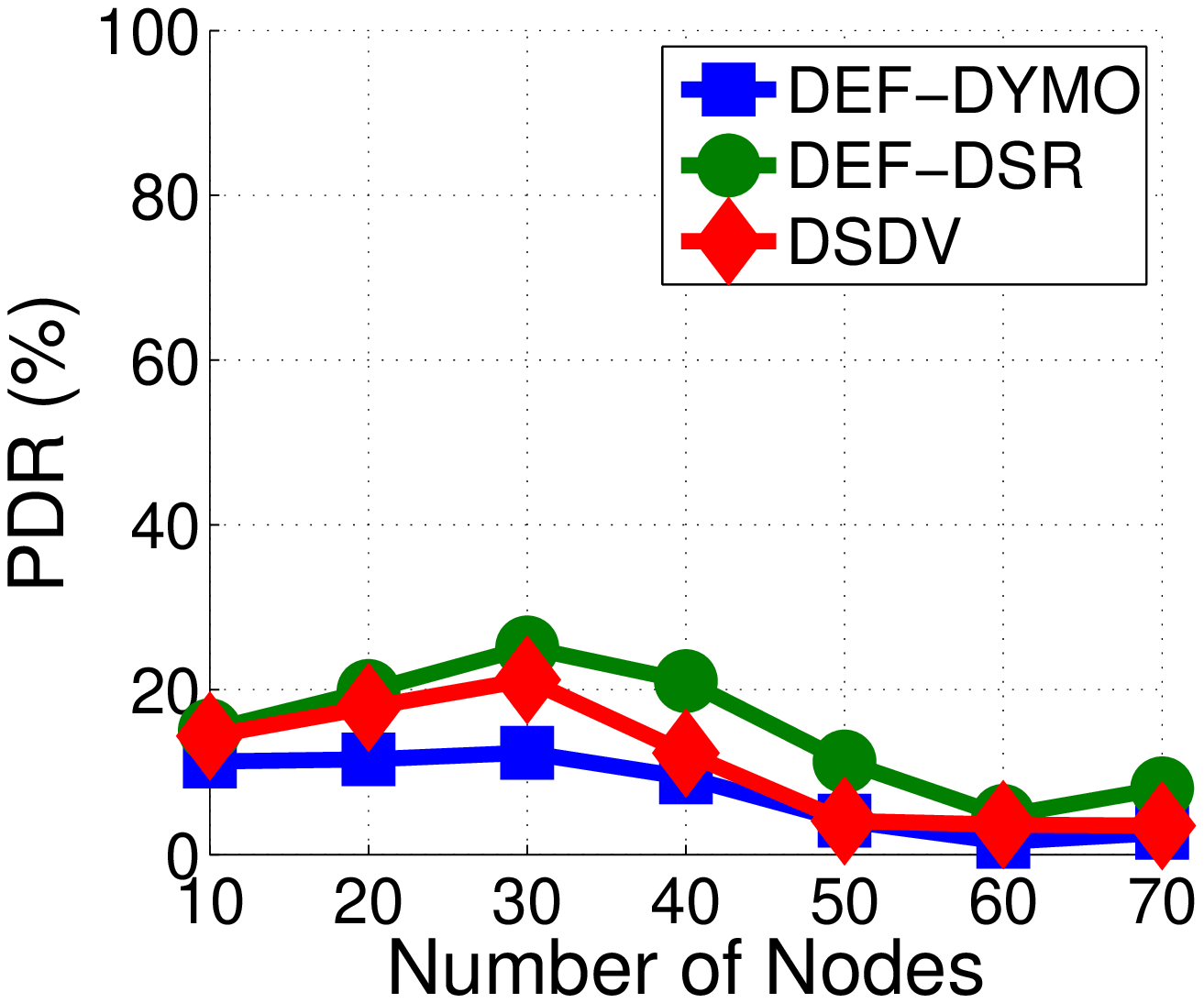}}
 \subfigure[PDR of Orig. Prot.s VANETs]{\includegraphics[height=2.7  cm,width=4.3 cm]{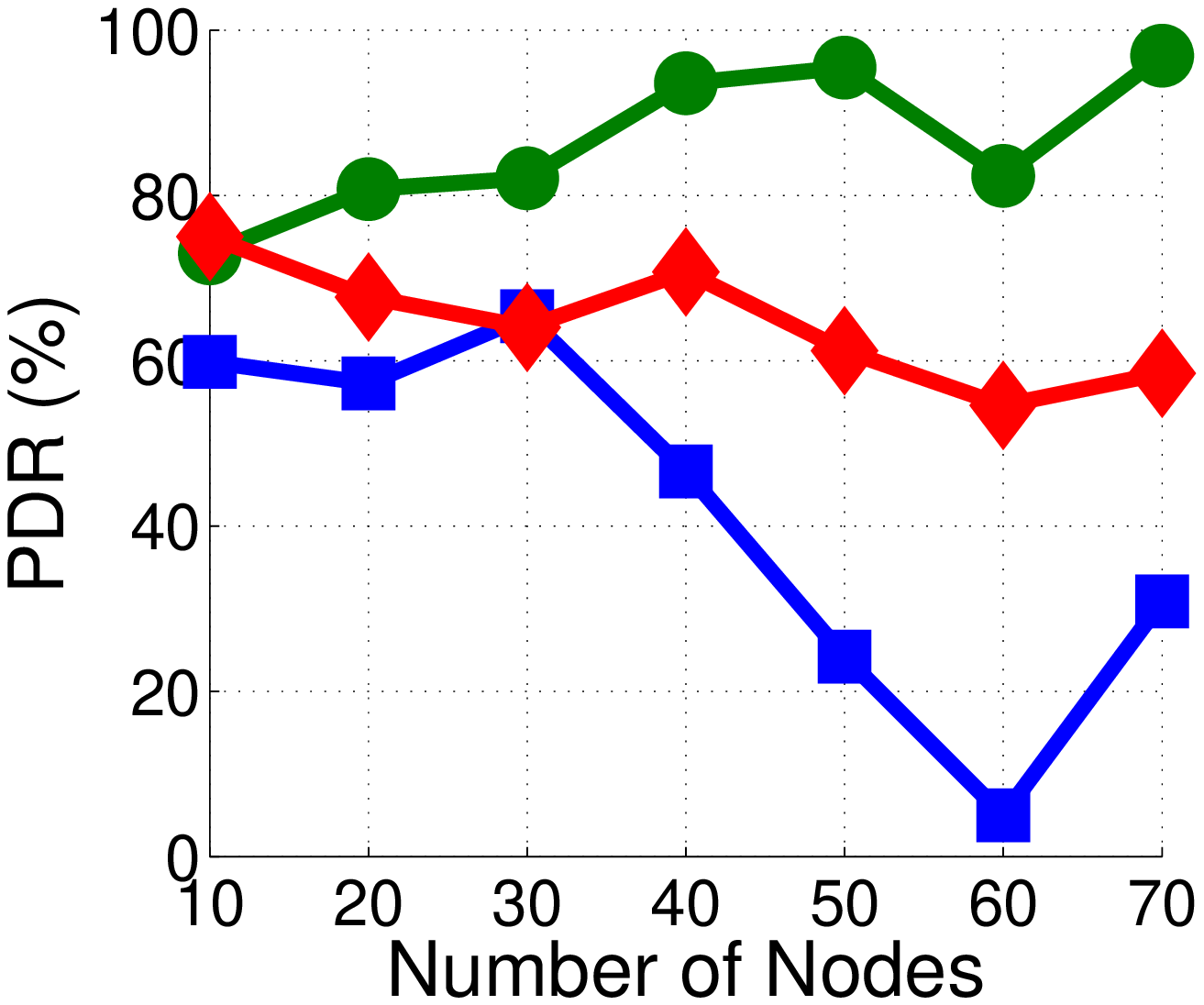}}
 \subfigure[PDR of Mod. Prot.s MANETs]{\includegraphics[height=2.7 cm,width=4.3 cm]{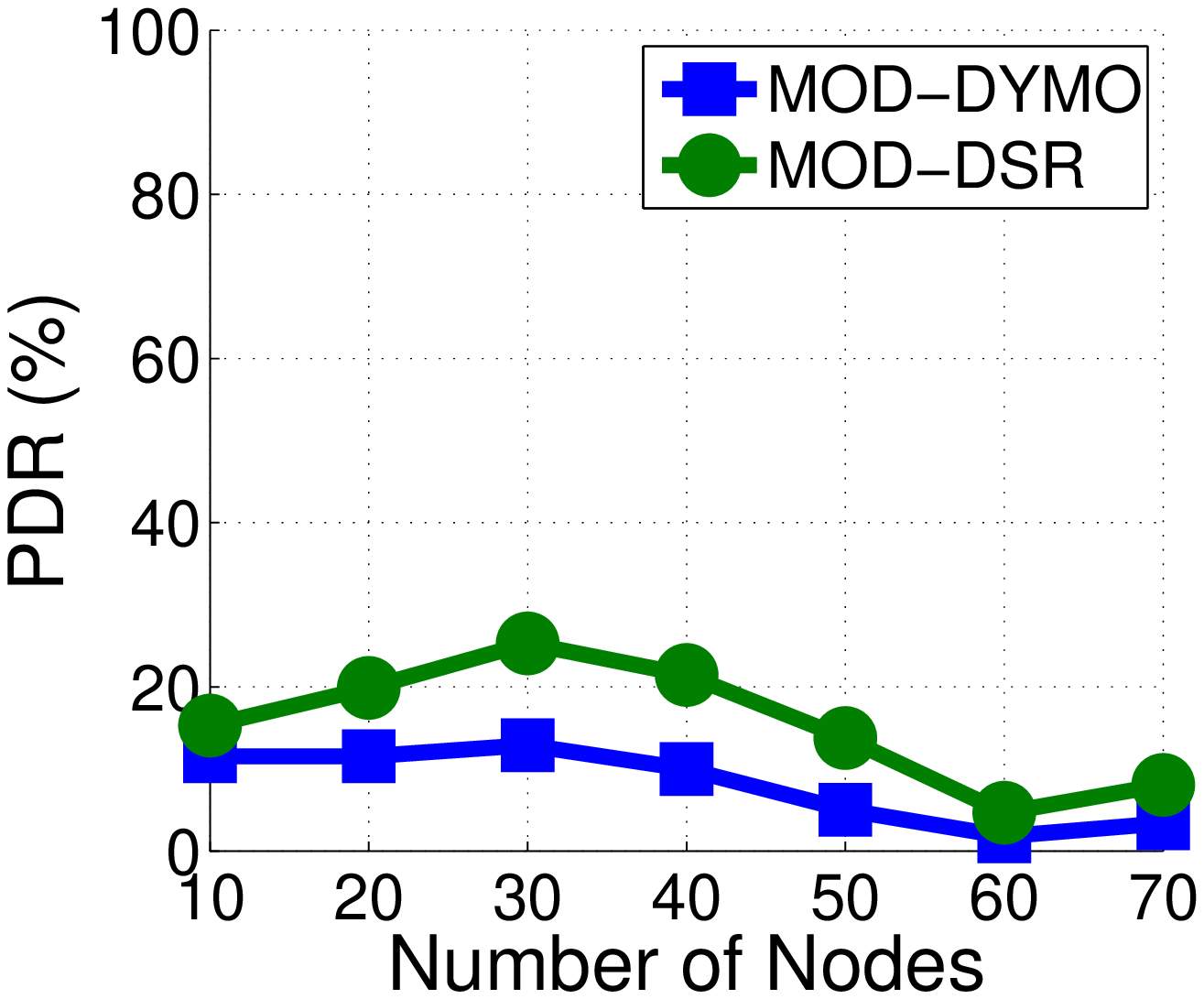}}
 \subfigure[PDR of Mod. Prot.s VANETs]{\includegraphics[height=2.7  cm,width=4.3 cm]{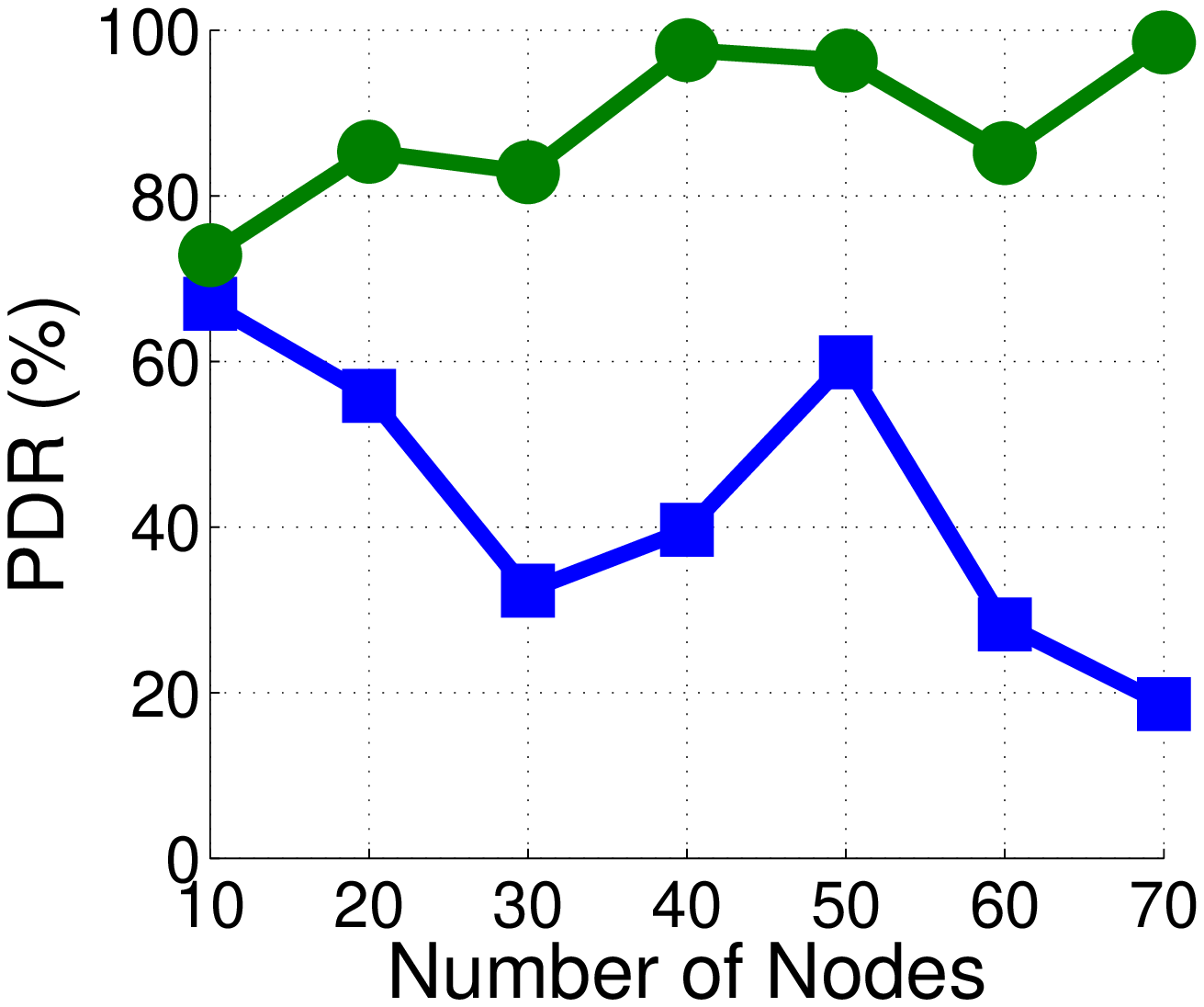}}
 \caption{PDR achieved by three routing protocols}
   \end{figure}

\subsection{E2ED}
Generally, DSR possesses the highest routing delay in both MANETs and VANETs. First checking of route cache for alternative routes requires more time as compared to simple ERS (as in DYMO) which causes delay in DSR. On the other hand, MOD-DSR due to frequently deletion of learned routes lessens packet slavaging  and route caching, as depicted in Fig. 6(d)(f) thus increase path latencies, as compared to MOD-DYMO. In general, E2ED of proactive protocol DSDV is lower as compared to reactive protocols both in MANETs and in VANETs because of pre-computation of routes.

\vspace{-0.2cm}
\subsection{NRO}
Among reactive protocols, DYMO attains the highest routing load among reactive protocols because of lack of any auxiliary mechanism, as depicted in Fig. 7. Whereas, DSDV attains lowest routing load in lower scalabilities and highest routing load in higher scalabilities. In MOD-DYMO control, packet generation becomes less due to increasing TTL values of ring thus, it has lower routing load in all scalabilities.

\begin{figure}[!h]
  \centering
 \subfigure[AE2ED of Orig. Prot.s MANETs]{\includegraphics[height=2.7 cm,width=4.3 cm]{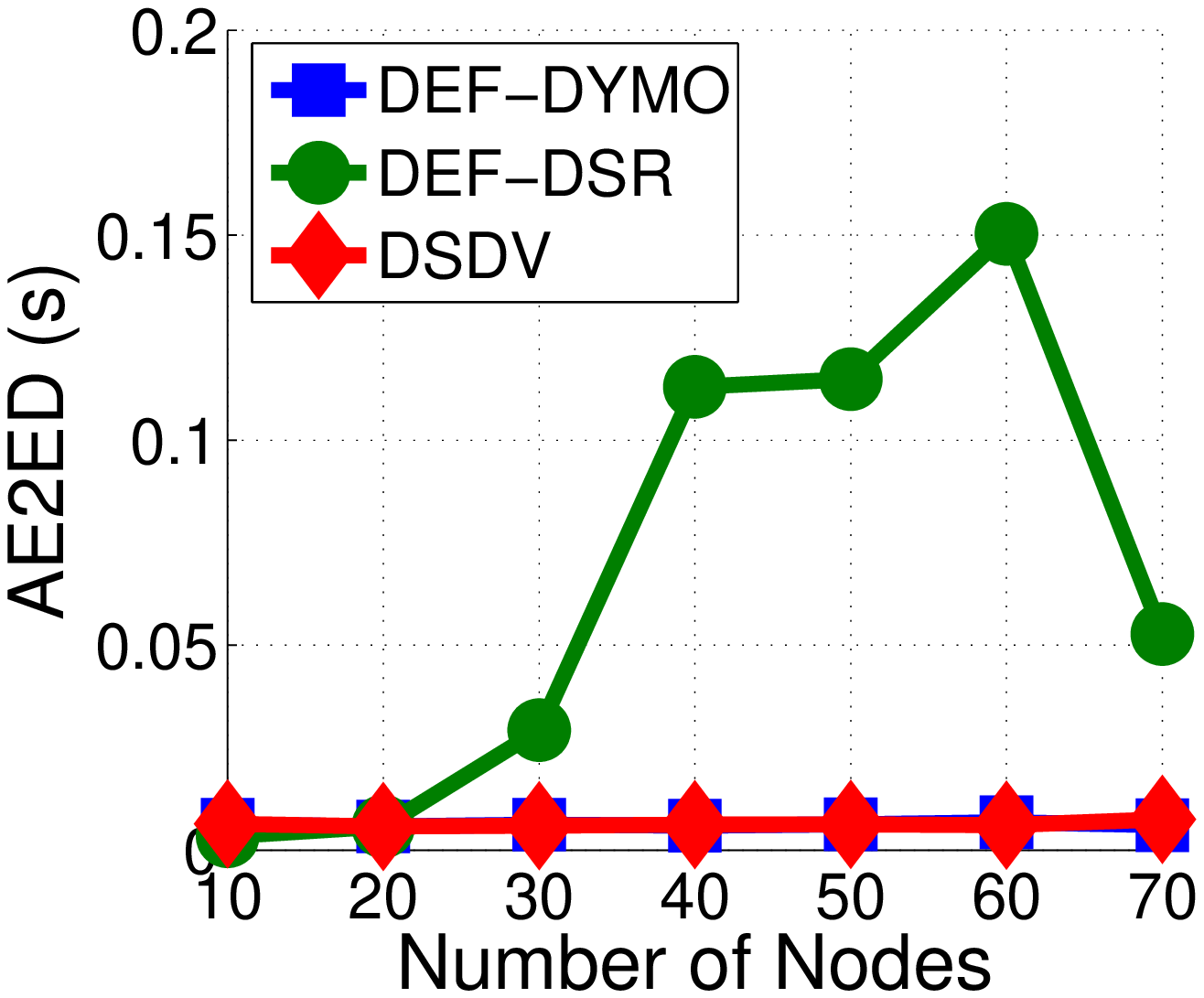}}
 \subfigure[AE2ED of Orig. Prot.s VANETs]{\includegraphics[height=2.7 cm,width=4.3 cm]{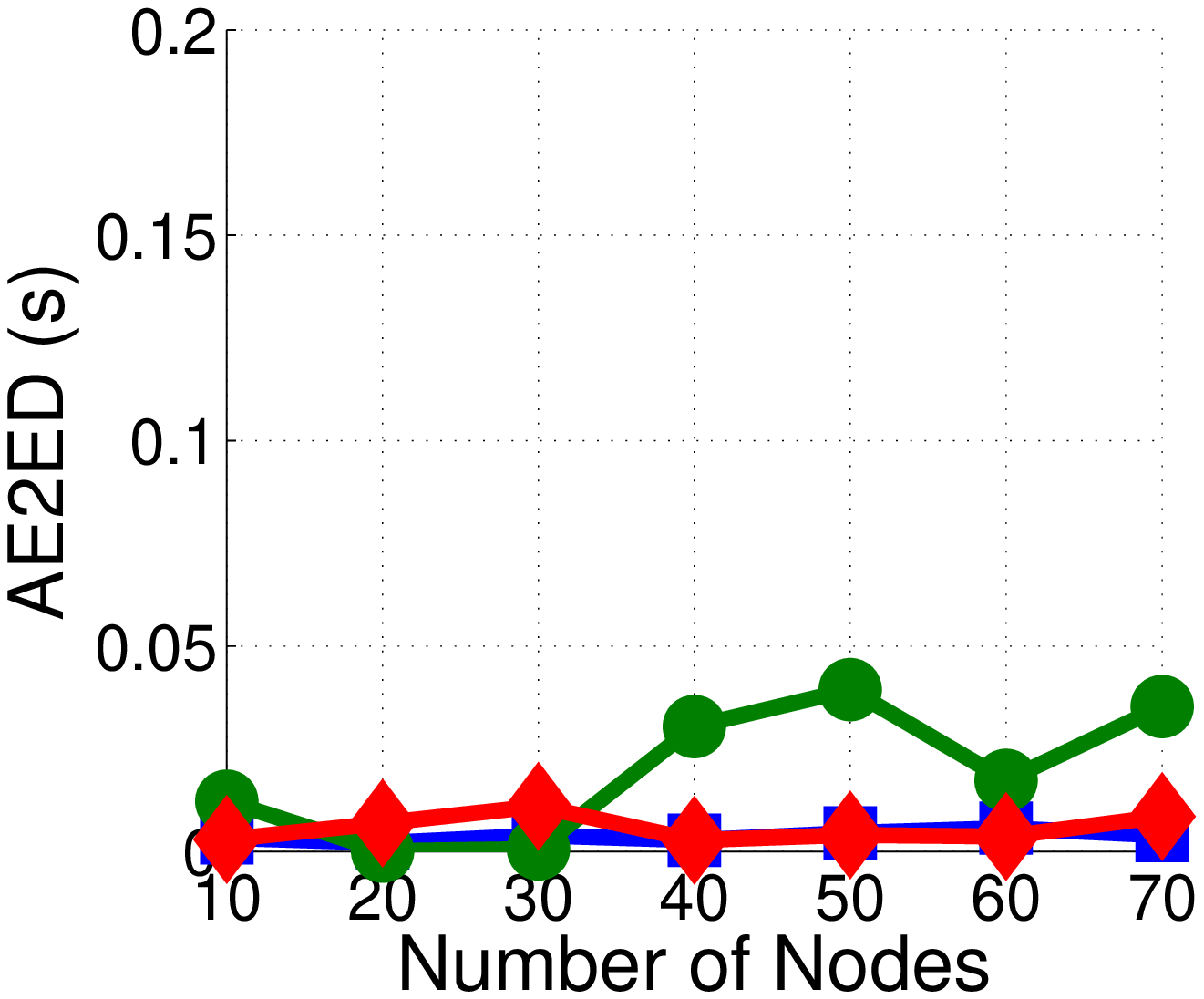}}
 \subfigure[AE2ED of Mod. Prot.s MANETs]{\includegraphics[height=2.7 cm,width=4.3 cm]{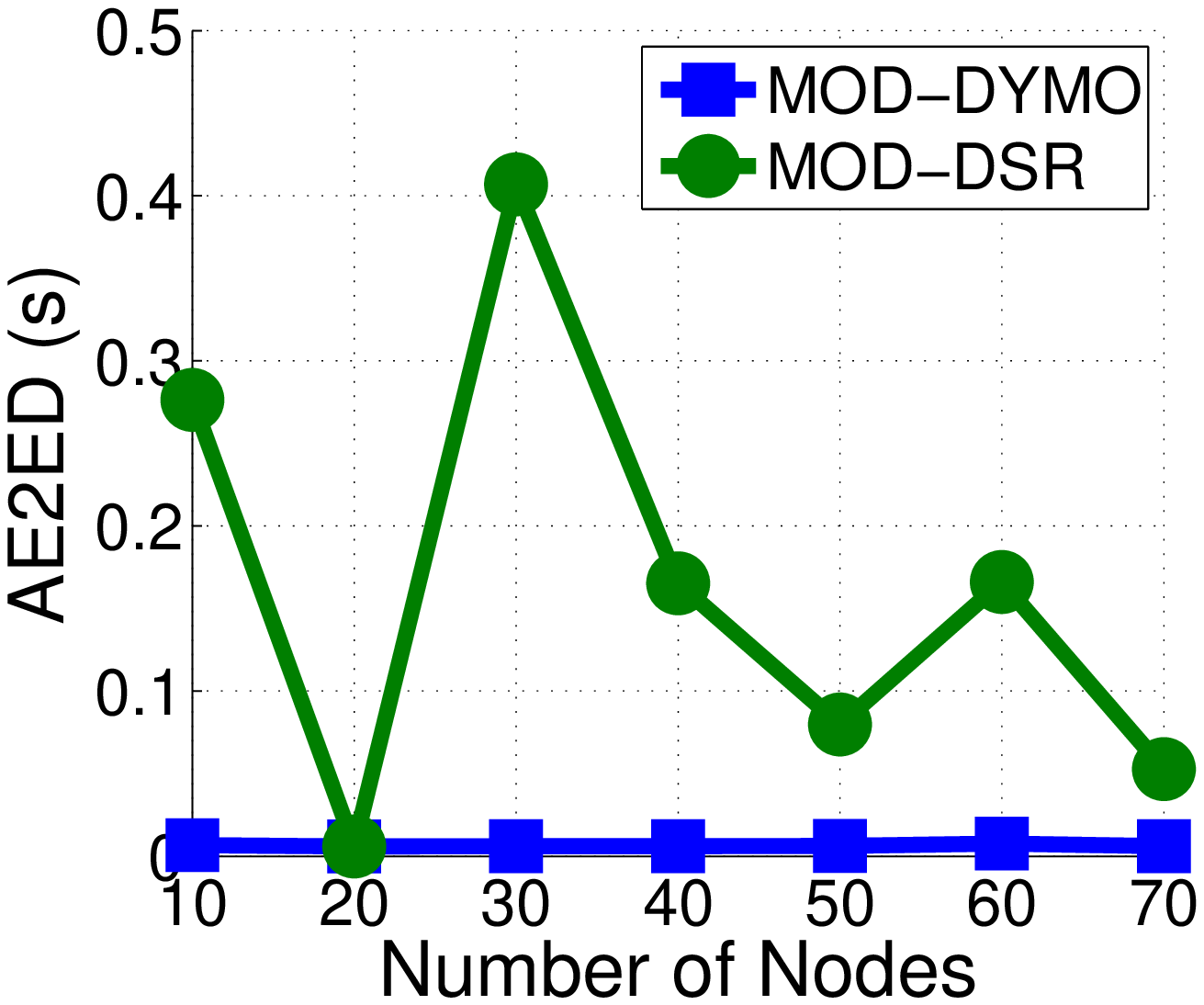}}
 \subfigure[AE2ED of Mod. Prot.s VANETs]{\includegraphics[height=2.7 cm,width=4.3 cm]{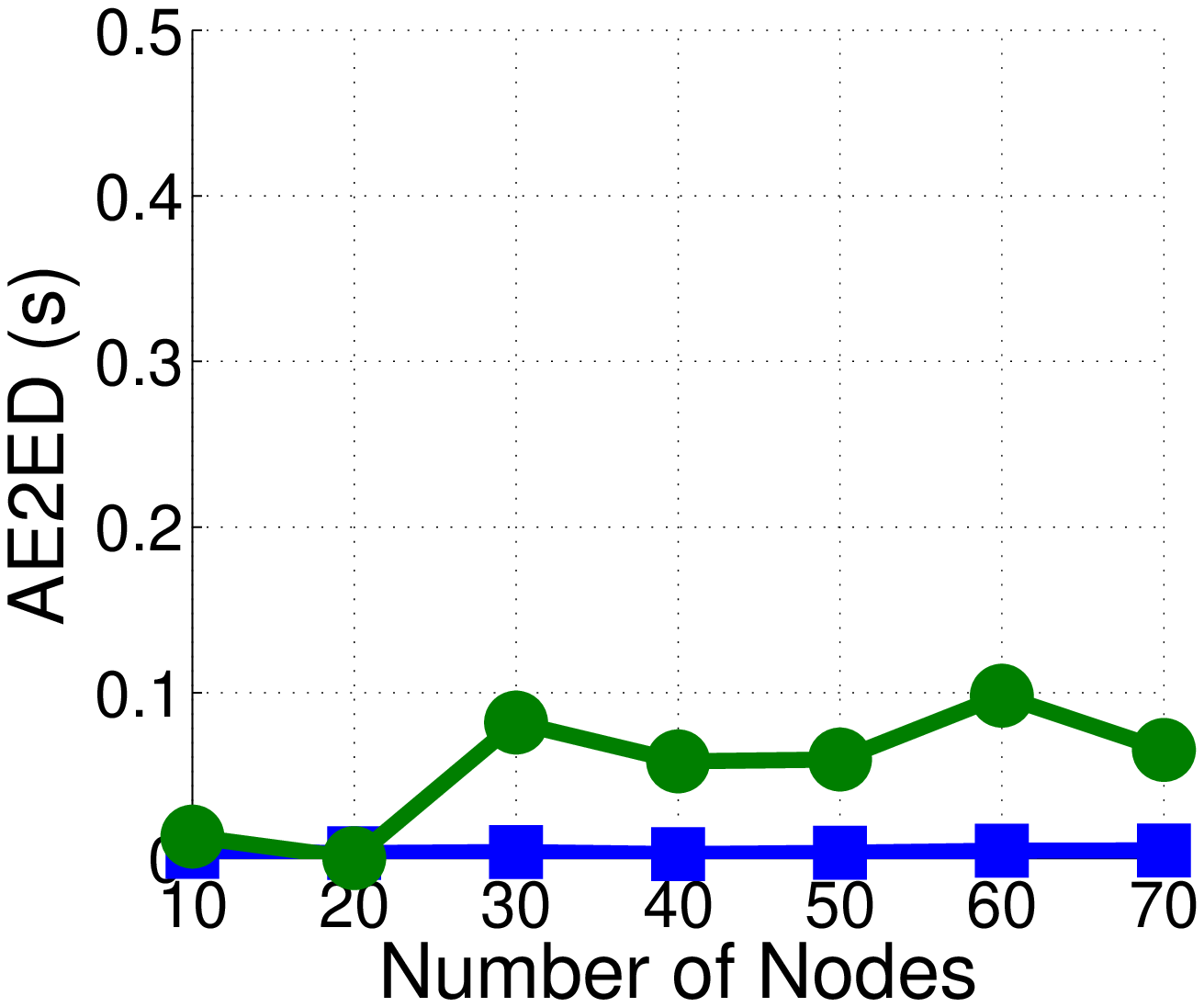}}
 \caption{End-to-end delay produced by protocols}
   \end{figure}

 \begin{figure}[!h]
  \centering
 \subfigure[NRO of Orig. Prot.s MANETs]{\includegraphics[height=2.7 cm,width=4.3 cm]{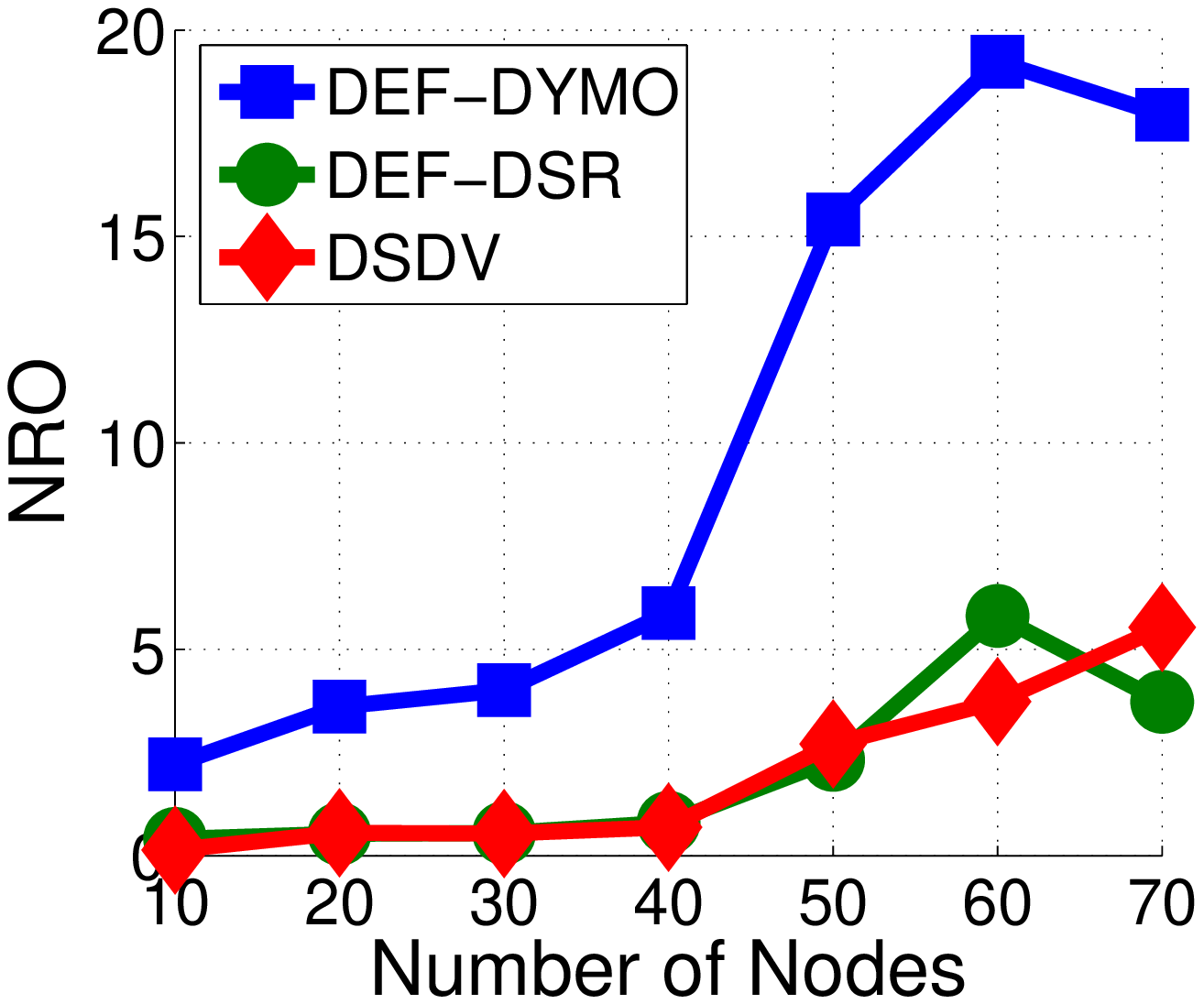}}
 \subfigure[NRO of Orig. Prot.s VANETs]{\includegraphics[height=2.7 cm,width=4.3 cm]{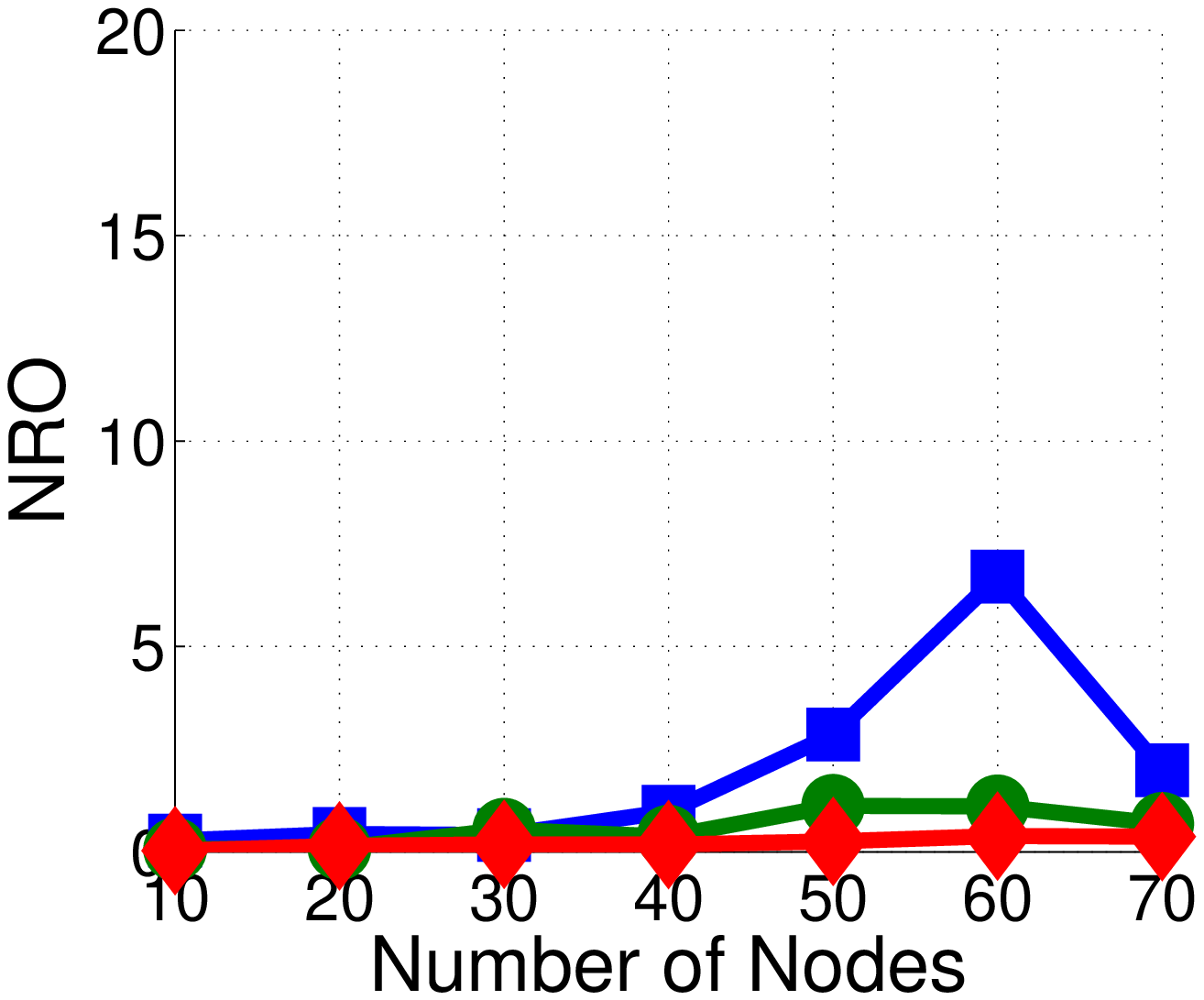}}
 \subfigure[NRO of Mod. Prot.s MANETs]{\includegraphics[height=2.7 cm,width=4.3 cm]{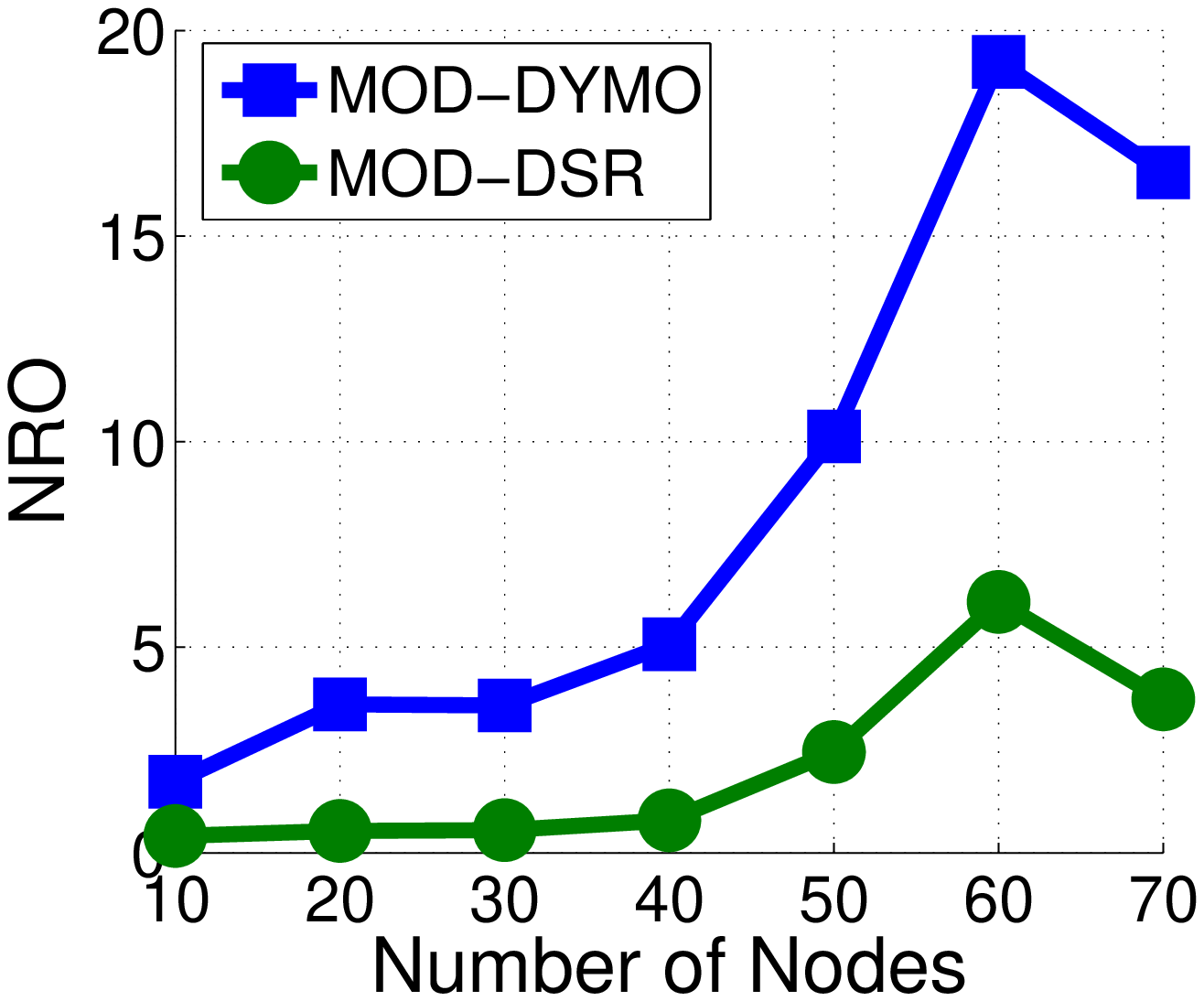}}
 \subfigure[NRO of Mod. Prot.s VANETs]{\includegraphics[height=2.7 cm,width=4.3 cm]{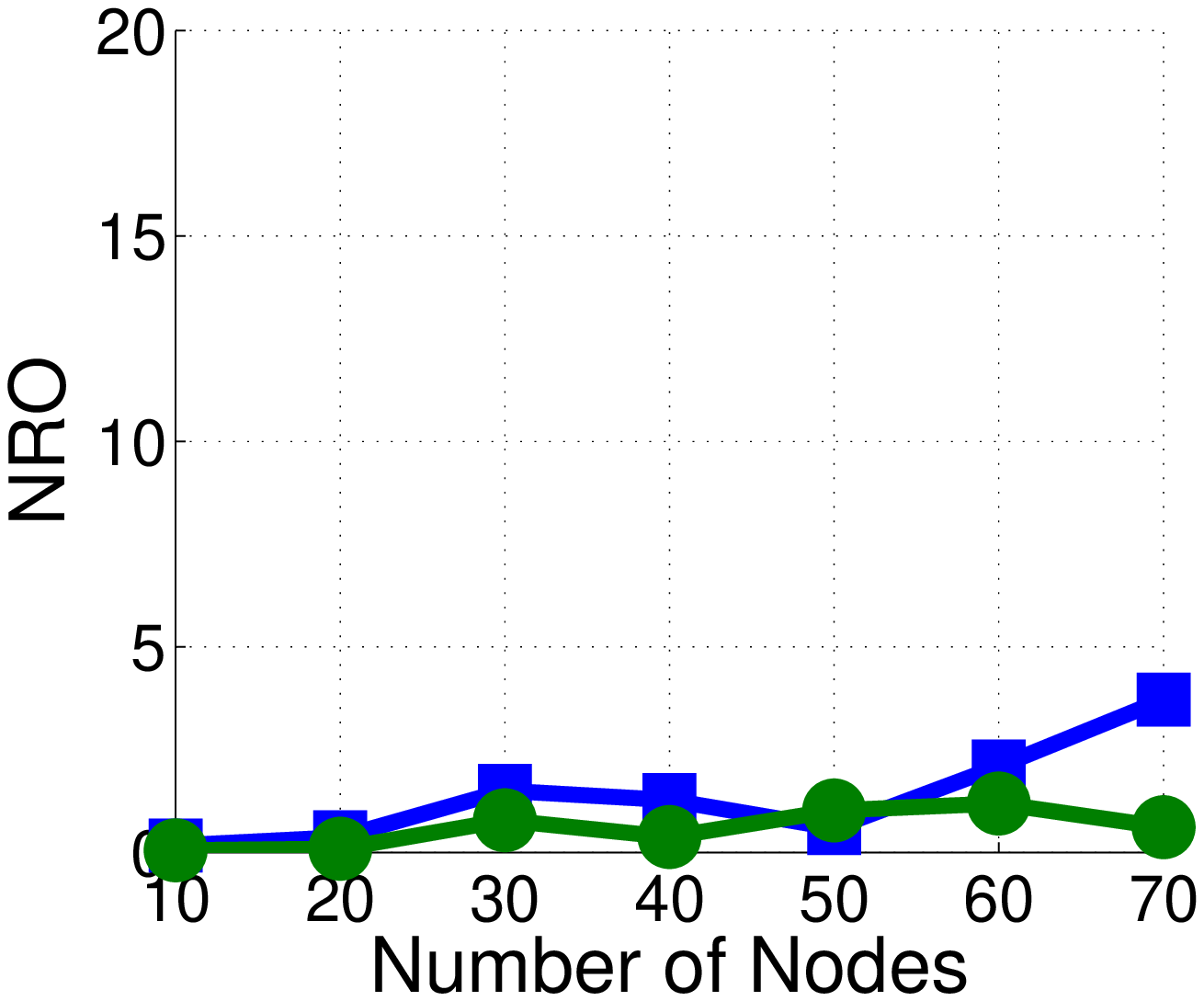}}
 \caption{Routing overhead faced by protocols}
   \end{figure}
\vspace{-0.2cm}
\section{Trade-offs Made by Routing Protocols to Achieve Performance}

DSDV: It attains low E2ED due to trigger updates, which provides instance convergence with correct route entry at the cost of NRO, because trigger updates flooded in entire network after detecting any link change in active routes (Fig. 6 as compared to Fig. 7).

DSR:  Grat. RREPs during RD due to route caching generate more routing load in DEF-DSR, as shown in Fig. 7, while these RREPs lower E2ED (in Fig. 6). On the other hand, in MOD-DSR, due to shortening $TAP\_CACHE\_SIZE$, E2ED is increased while PDR is also increased in VANETs.

DYMO: Simple ERS without any supplementary mechanism like grat. RREPs in DEF-DYMO augments NRO, whereas its E2ED is less as compared to DSR, as can be seen from Fig. 6(a)(b) comparative to Fig. 7(a)(b). While, increasing the TTL values of ERS for MOD-DYMO then NRO decreases in MANETs as well as in VANETs, as shown in Fig. 7(c)(d). Whereas, this modification increased E2ED as depicted in Fig. 6(c)(d).
\vspace{-0.2cm}
\section{Conclusion}
\vspace{-0.1cm}
In this paper, a framework for experiment performance parameters, PDR, E2ED and NRO is presented for DSR, DYMO and DSDV and also their theoretical results are analyzed with siome assumptions. A novel contribution of this work is enhancement of DSR and DYMO protocol to improve performance efficiency in VANETs. Moreover, we also evaluate the protocols in MANETs and in VANETs using NS-2
simulator and TwoRayGround radio propagation model. The SUMO simulator is used to generate mobility pattern for VANET to evaluate the performance of selected routing protocols for three performance parameters i.e., PDR, E2ED and NRO. Our simulation results show that DSR performs better at the cost of delay both in MANETs and in VANETs. In future, we are interested to develop a new link metric, like [16] and [17].

\end{document}